\documentclass[11pt]{article}

\usepackage{graphicx}
\begin{document}
\title{\bf Young Open Star Clusters: The Spatial Distribution of Metallicity
in the Solar Neighborhood}

\author{{M.\,L.~Gozha,  V.\,A.~Marsakov} \\
{Southern Federal University, Rostov-on-Don, Russia}\\
{e-mail: gozhamarina@mail.ru,  marsakov@ip.rsu.ru }}
\date{accepted \ 2013, Astronomy Letters, Vol. 39, No. 3, pp. 171-178}

\maketitle

\begin{abstract}

We perform a comparative analysis of the spatial distribution 
of young ($ < 50$~Myr) open star clusters and field Cepheids with 
different metallicities. A significant fraction of young clusters 
are shown to have low metallicities atypical of field Cepheids. 
Both types of objects exhibit approximately equal (in magnitude) 
negative radial metallicity gradients, while their azimuthal 
metallicity gradients differ outside the error limits and have 
opposite signs. Among the stellar complexes identified by young 
clusters, the most metal--poor clusters are grouped in the Perseus 
complex. It is the clusters of this complex that are responsible 
for the radial and azimuthal metallicity gradients among young 
clusters. The described properties are indicative of a weak mixing 
of interstellar matter before the onset of star formation there.
Significant differences between the spatial distributions of open 
clusters and field stars with different metallicities suggest different 
conditions required for the formation of these types of objects.

\end{abstract}

{\em Keywords:}open star clusters, field Cepheids, stellar complexes, 
chemical composition, Galaxy.

\maketitle

\section{Introduction}

Previously (Gozha et\,al. 2012a,~2012b), we
showed that the population of open star clusters is
heterogeneous and is divided into two groups differing
by their mean parameters, properties, and origin. The
first group includes the Galactic clusters that were
formed mainly from the interstellar matter of the thin
disk, have nearly solar metallicities ($[Fe/H] > -0.2$),
and possess almost circular orbits a short distance
away from the Galactic plane, i.\,e., typical of the field
stars of the galactic thin disk. The second group
includes the peculiar clusters formed through the
interaction of extragalactic objects (such as highvelocity
clouds, globular clusters, or dwarf galaxies)
with the interstellar matter of the thin disk,
which, as a result, derived abnormally low (for field
thin--disk stars) metallicities and/or Galactic orbits
typical of objects of the older Galactic subsystems.
Analyzing the orbital elements, we also showed in
the above papers that the bulk of the clusters from
both groups were formed within a Galactocentric
radius of $\approx 10.5$~kpc and closer than $ \approx 180$~pc from
the Galactic plane, but owing to their high initial
velocities, the peculiar clusters gradually took up the
volumes occupied by the objects of the thick disk, the
halo, and even the accreted halo (i.\,e., the corona) of
the Galaxy.

In this paper, we are interested in the spatial distribution
of heavy elements in the interstellar medium
in the solar neighborhood before the onset of star
formation there. Open star clusters are perhaps the
most suitable objects for such a study. Several factors
contribute to this. First, clusters are visible at considerable
distances, and these distances are determined
fairly accurately. Second, the ages of clusters are
determined much more reliably than those of isolated
field stars. Third, among the clusters, there are many
very young ones that could not go far away from their
birthplaces because of their young ages. That is why
they clearly trace the segments of spiral arms that
are giant stellar complexes. And, finally, fourth, the
metallicities can often be determined for F–G stars
of even very distant clusters not only by photometric
methods but also by spectroscopic ones.

Short--lived long--period Cepheids are equally good
tracers of spiral arms and typical representatives of
stellar complexes. Since the metallicities are known
for a considerable number of these objects, the spatial
metallicity distributions for field stars and clusters
can be compared. Based on Cepheids, Lepine et\,al.
~(2011) detected an azimuthalmetallicity gradient that
turned out to be comparable in magnitude the radial
one. This means that the metallicities of field
stars located at the same Galactocentric distances
exhibit appreciable systematic variations of the chemical
composition, indicative of a weak mixing of the
interstellar medium. As a result, these authors conclude
that the radial metallicity gradient in the thin
disk detected by many objects in the solar neighborhood
points only to the existence of a statistical
trend, and its values can differ in different directions
from the Galactic center. The goal of this paper is to
perform a comparative analysis of the spatial distributions
of heavy elements in the solar neighborhood
among young open clusters and field Cepheids and to
investigate properties of clusters inside giant stellar
complexes.


\section {Initial data}

Previously (Gozha et\,al.~2012a), we described
the catalog of fundamental astrophysical parameters
that we compiled based on the most recent
published data for 593 Galactic open clusters with
known total velocities or metallicities1(The catalog
is accessible in electronic form at http://vizier.ustrasbg.
fr/cats/J.PAZh.htx). This catalog contains
226 clusters younger than 50\,Myr; the distances are
known for all of them, and the masses and metallicities
are known for 192 and 57 clusters, respectively.
Since the overwhelming majority of sample clusters
are within 3\,kpc of the Sun, these parameters were
determined for them with an accuracy sufficient for
statistical studies. The errors of all the parameters
used are described in detail in the paper cited above.

A list containing 276~field Cepheids with known
variability periods, distances, and space velocity components
was taken from Berdnikov et\,al.~(2003). We
calculated the ages of these stars from the formula
$log t = 8.16 - 0.68 log P$, where $t$ is the age in years
and $P$ is the Cepheid period in days. The list contains
207 Cepheids younger than 50\,Myr. Spectroscopic
determinations of the iron abundances for 77 young
Cepheids were found in the papers of one group of
authors (Andrievsky et\,al.~2002a.2002c, 2004; Luck
et\,al.~2003; Kovtyukh et\,al.~2005).

Recall that in our previous papers (Gozha et\,al.
~2012a, 2012b) we determined the membership of
stars and clusters in a particular Galactic subsystem
from the Galactic orbital elements. As a generalized
orbital characteristic, we used the parameter proposed
by Chiappini et\,al. (1997), $(Z^{2}_{max} + 4e^{2})^{1/2}$,
where the eccentricity (e) is a dimensionless quantity
and the maximum distance of the orbital points from
the Galactic plane ($Z_{max}$) is measured in kiloparsecs.
Figure 1a shows the age --- $(Z^2_{max} + 4e^2)^{1/2}$ diagram.
It can be seen from this diagram that almost all of
the sample clusters with high eccentric orbits (i.\,e., by
definition from Gozha et\,al.~(2012a), those satisfying
the criterion ($Z^2_{max} + 4e^2)^{1/2} > 0.35$) turned out to
be younger than 20 Myr. This implies that the last
formation burst of such ЃgfastЃh clusters began precisely
20~Myr ago. Unfortunately, the ratios [Fe/H]
are known only for four of the 27~young fast clusters
(see Fig.~1b), and all of them have a nearly solar
metallicity. According to Vande Putte et\,al.~(2010),
such clusters were formed due to globular cluster
impact on the Galactic disk, but, on the other hand,
almost all metal--poor ($[Fe/H] < \approx 0.25$) young
clusters are in flat, nearly circular orbits. According to
Vande Putte et\,al.~(2010), such clusters were formed
from the interstellar matter that fell from the outer
parts of the Galaxy (or was captured from disrupted
companion galaxies).

Analysis shows that the most probable age of our young clusters 
is $ \approx 10$~Myr (see Fig.~2b), while their most probable  velocity 
relative Local Standard of Rest is $V_{LSR} \approx 18$~km\,s$^{-1}$  
(see Fig.~1c, where the five fastest clusters are not shown).
As a result, the mean displacement of the clusters
relative to their birthplaces turned out to be less than
200 pc, i.\,e., approximately of the same order of magnitude
as the error in the distances to these clusters,
$ \approx 20\,\%$ (see Gozha et\,al.~2012a). This implies that
clusters are quite suitable for a statistical analysis
of the distribution of chemical elements in the solar
neighborhood. Since these displacements were found
to be several times larger for fast clusters, they are
already far from their birthplaces even for a young age.

Figure~1d shows the distribution of young clusters
in difference between the current position of the cluster
($R_{G}$) and the mean radius of its $R_{m}$ normalized
to ($R_{a} - R_{p}$). (In view of the fact that the orbital
parameters we use were calculated by Vande Putte
et\,al.~(2010) as the means over 15~Gyr, the current
positions of several clusters were found to be less
than the tabulated perigalactic radii ($R_{p}$) or greater
than the apogalactic radii (Ra) of their orbits, i.\,e.,
the ratios $|(R_{G} - R_{m})/(R_{a} - R_{p})| > 0.5$ for them.)
Since the ages of our clusters are much less than
their revolution periods around the Galactic center,
the part of the orbit where the cluster was formed can
be determined from the histogram. It can be seen
from the diagram that the clusters are formed predominantly
near the apogalactic radii of their orbits
(the ratios are $> 0.4$ for one third of the clusters),
more rarely near the perigalactic radii, and even more
rarely near themean orbital radii. Thismanifests itself
to an even greater extent for fast clusters with high
eccentric orbits (the gray histogram in Fig.~1d). Such
clusters are formed mainly near the maximum radii of
their orbits (the ratios are $> 0.4$ for more than half of
the clusters). This means that they acquire an initial
momentumfrom extragalactic objects moving toward
the Galactic center.

{\bf \em COMPARATIVE ANALYSIS
OF THE PROPERTIES OF YOUNG OPEN
CLUSTERS AND FIELD CEPHEIDS}.

Previously (Gozha et\,al.~2012a), we showed that
the metallicity distributions for open clusters and
nearby thin-disk field F-–G stars of all ages differ
significantly. Let us now compare the distributions
for objects at approximately equal distances from the
Sun-—young clusters and field Cepheids. Figure~2a
presents the metallicity functions for these objects.
There is an almost complete coincidence of the distribution
features with the histograms in Fig.~2c from
Gozha et\,al.~(2012a). In particular, a clear excess
of metal--poor clusters compared to field Cepheids is
observed. Similarly, the main maximum of the [Fe/H]
distribution for young clusters has a metallicity larger
than the solar one, while for field Cepheids it is slightly
smaller than the solar one.

Although the objects of both groups are young,
their age distributions differ: the maximum for clusters
is observed near 10~Myr, while for Cepheids it
is later approximately by 30~Myr. This is because a
massive star takes some time to evolve to the pulsational
instability strip and to become a Cepheid. On
the other hand, young clusters experience a strong
gravitational influence from the spiral density wave
that formed them and, therefore, dissipate quickly.
As a result, the relative number of clusters decreases
sharply already after 20~Myr. Nevertheless, the number
of objects of both types within our chosen range
($< 50$~Myr) turns out to be approximately the same
and is sufficient for statistically significant estimations.

For both types of objects, Fig.~3a presents the
age-–metallicity relations within 50~Myr. We see that
a significant fraction of young clusters also turn out
to be below the lower envelope for field Cepheids
(see the dashed curve in the diagram drawn by eye).
This provides convincing evidence for both significant
inhomogeneity of the chemical composition of
the interstellar medium in the Galactic plane in the
solar neighborhood and a different history of chemical
evolution of the matter from which the metal-poor
clusters were formed. Although the Galactocentric
distance–metallicity diagrams (Fig. 3b) presented on
the next panel demonstrate approximately identical,
within the error limits, radial metallicity gradients for
both groups of objects, the relation for clusters lies,
as would be expected, well below that for field stars.
In contrast, the azimuthal gradients for field stars
and clusters differ radically (see Fig.~3c, where the y
coordinate is positive in the direction of Galactic rotation).
Whereas young Cepheids exhibit an increase in
metallicity in the direction of Galactic rotation, confirming
the conclusions by Lepine et\,al.~(2011), young
clusters exhibit a decrease. The negative azimuthal
metallicity gradient for clusters is provided exclusively
by the objects lying below the lower envelope for field
Cepheids on the age–metallicity diagram (Fig.~3a).
If these clusters are removed from the diagram, then
the relations for the remainingmetal--rich clusters and
field Cepheids will coincide completely. If the same
clusters are removed from the previous panel, then the
radial gradient for young clusters will become zero,
within the error limits.

{\bf SPIRAL ARMS AND STELLAR COMPLEXES}. 

It is generally believed that star formation in the
thin Galactic disk is stimulated mainly by spiral density
waves. According to present views, the segments
of spiral arms detected in the solar neighborhood are
giant stellar complexes (Efremov~2011). Young star
clusters and field Cepheids confirm this with confidence.
Figure 4a presents the distributions of young
(younger than 50 Myr) open clusters and equally
young field Cepheids in projection onto the Galactic
plane. It can be seen from the figure that although
the formation of both types objects is assumed to be
stimulated in a common process, the regions with enhanced
densities for them do not coincide. Arbitrary
boundaries between these regions (see the curves on
the diagram drawn by eye) that are located along the
usually identified spiral arms can even be drawn. Although
the objects being studied are young and went
not far away from their birthplaces, this separation
does not necessarily suggest that the clusters and
field stars were born independently of one another.
Recall that the mean age of the Cepheids in our sample
($33 \pm 11$~Myr) is greater than that for the sample
of young clusters ($17 \pm 12$~Myr). As a result, for
$t < 50$~Myr, 70\,\% of the clusters turn out to be younger
than 20 Myr, while the fraction of such objects among
the Cepheids is only 12\,\%. Some authors (see, e.\,g.,
Ivanov~1983) point out that, for example, the clusters
on the inner wing of the Carina–-Sagittarius arm are
systematically younger than those at the outer edge,
as is required by the wave theory of spiral structure.
The difference between the spatial distributions of
clusters and field stars observed in Fig.~4a is most
likely also due to the wave nature of the spiral structure,
when conditions favorable for the formation of
clusters and field stars are created at the collision
front of the interstellar medium overtaking the density
wave. After several tens of Myr, the number
of clusters at the outer edge of the wave decreases
considerably, while the Cepheids dominate for such
an age.

However, as has already been pointed out, the
clusters and field Cepheids exhibit a significant difference
in heavy--element abundances. It can be seen
from Fig. 4a that the relative number of metal-poor
Cepheids on the diagram (the crosses on the diagram
are Cepheids with $[Fe/H]< 0.0$) gradually decreases
from left to right and from bottom to top. On the
other hand, the metal-poor clusters (i.\,e., those lying
below the lower envelope for fieldCepheids in Fig.~3a),
with amore or less uniformdistribution over the entire
diagram, form a clump in its upper left corner (see the
closed circles in Fig.~4b). It is this clump (which is the
Perseus stellar complex, see below) that is responsible
for the negative radial and azimuthal metallicity
gradients for young clusters (on the panels of Fig.~3,
the clusters of the Perseus complex are indicated by
the filled circles). The metal--rich clusters also form
a clump but at the center of the diagram (the open
circles in Fig.~4b). In other words, the spatial distributions
of young open clusters with different metallicities
are very nonuniform. Occasionally, clusters
of different metallicities are close neighbors within
the same complex (see the filled and open circles in
Fig.~4b).

The ovals drawn by eye in Fig.~4b indicate five
regions with enhanced cluster densities usually associated
with the stellar complexes that are the segments
of the Perseus, Cygnus, Sagittarius, Carina,
and Local-system spiral arms. The distributions of
most physical parameters for the identified groups
of clusters differ from one another statistically insignificantly,
but there are also some differences. In
particular, the clusters of the Local system clearly
show a bimodal age distribution with a rather narrow
dominant peak at $t = 11 \pm 1$ Myr including two thirds
of the clusters and a very distant lower peak at $t =
37 \pm 1$~Myr (Fig.~5a). As can be seen from the same
figure, the age distribution for all clusters exhibits a
sharp decrease in the number to 20~Myr years and
then the decrease continues much more slowly. The
mean mass of the Local-system clusters also turned
out to be lower, outside the error limits, than themean
over all young clusters (see Fig.~5b). Themetallicities
for all (except one) clusters of the Local system are as
high as those for Cepheids.

Another statistically significant feature is a radical
difference between the metallicity distributions for the
clusters of the Perseus complex and all of the remaining
young clusters (see Fig.~5c). These clusters are
the most metal--poor ones among the young clusters,
and their orbits are almost flat and circular.

The Perseus group also contains six fast clusters
(the metallicity was determined only for one of them
and it is solar); they are distributed mainly along the
periphery of the complex (see the asterisks in Fig.~3b),
while seven of the twelve metal--poor clusters form
a dense core in the middle of the group. Five fast
clusters are also observed in the Carina complex, with
three of them lying near its boundary. There are no
clusters with high eccentric orbits in the Cygnus and
Sagittarius complexes at all, and only two are present
in the Local system, while the remaining (about fifteen)
fast young clusters are scattered over the entire
field of the diagram. Such a chaotic distribution of
fast clusters suggests that exclusively spiral density
waves were unlikely to be responsible for their formation.

{\bf\em CONCLUSIONS } 

Thus, we see that the population of young open
clusters is heterogeneous. In particular, some of them
have exhibited very large space velocities and high
eccentric orbits since the earliest age (see Fig.~1a).
This unequivocally points to a nonrelaxational nature
of these velocities. In addition, some of young clusters
have low metallicities atypical of thin-disk field stars
(Fig.~3a), although the orbits of such clusters are
flat and almost circular. Such a low metallicity is
most naturally explained by the fall of matter with a
different history of chemical evolution on the Galactic
disk and the predominant formation of precisely
open clusters rather than field stars from this matter.
This fall (from the outer regions of the Galaxy or from disrupted dwarf 
companion
galaxies) occurred not too long ago, because the interstellar
matter did not have time to be sufficiently
mixed before the starburst in it. In addition, both
young clusters and field Cepheids exhibit a chemical
composition inhomogeneity, when objects of the
same type with different heavy--element abundances
coexist at the same place (Fig.~4). A large--scale
inhomogeneity of objects in chemical composition is
also observed, when the clusters and Cepheids show
opposite (in sign) azimuthalmetallicity gradients that
are comparable in absolute value to the approximately
equal (in magnitude) negative radial metallicity gradients
exhibited by both types of objects. For young
clusters, both gradients are attributable exclusively
to the existence of the metal--poor Perseus complex,
while for field Cepheids, they are due to a gradual
increase in the relative number of metal-rich stars in
the directions of Galactic rotation and the Galactic
center. Note that both types of objects are good tracers
of spiral arms, suggesting the existence of a global
mechanism triggering star formation, spiral density
waves. The observed displacement of the clusters and
Cepheids, respectively, to the inner and outer edges
of the arms is most likely due to the slightly greater
mean age of the Cepheids and the wave nature of the
spiral structure. Thus, the described properties are
indicative of a difference between the conditions in the
interstellar medium required for the formation of open
clusters and field stars.
To extend our views of the degree of inhomogeneity
of the interstellar matter, the history of its chemical
evolution, and the origin of open star clusters, data on
the abundances of the chemical elements synthesized
in various processes in the atmospheres of open-cluster
stars are needed. Unfortunately, a detailed
chemical composition has been determined at present
only for about 60 open clusters and only for one young
cluster.

\section*{ACKNOWLEDGMENTS}

This work was supported  by the Russian Foundation for Basic 
Research (project no. 11-02-00621a), also with partial support 
from the Ministry of Education and science of the RF 
(projects no. P 685 and no. 14. A18.21.0787).

\renewcommand{\refname}{Список литературы}

\newpage

\begin{figure*}
\centering
\includegraphics[angle=0,width=0.96\textwidth,clip]{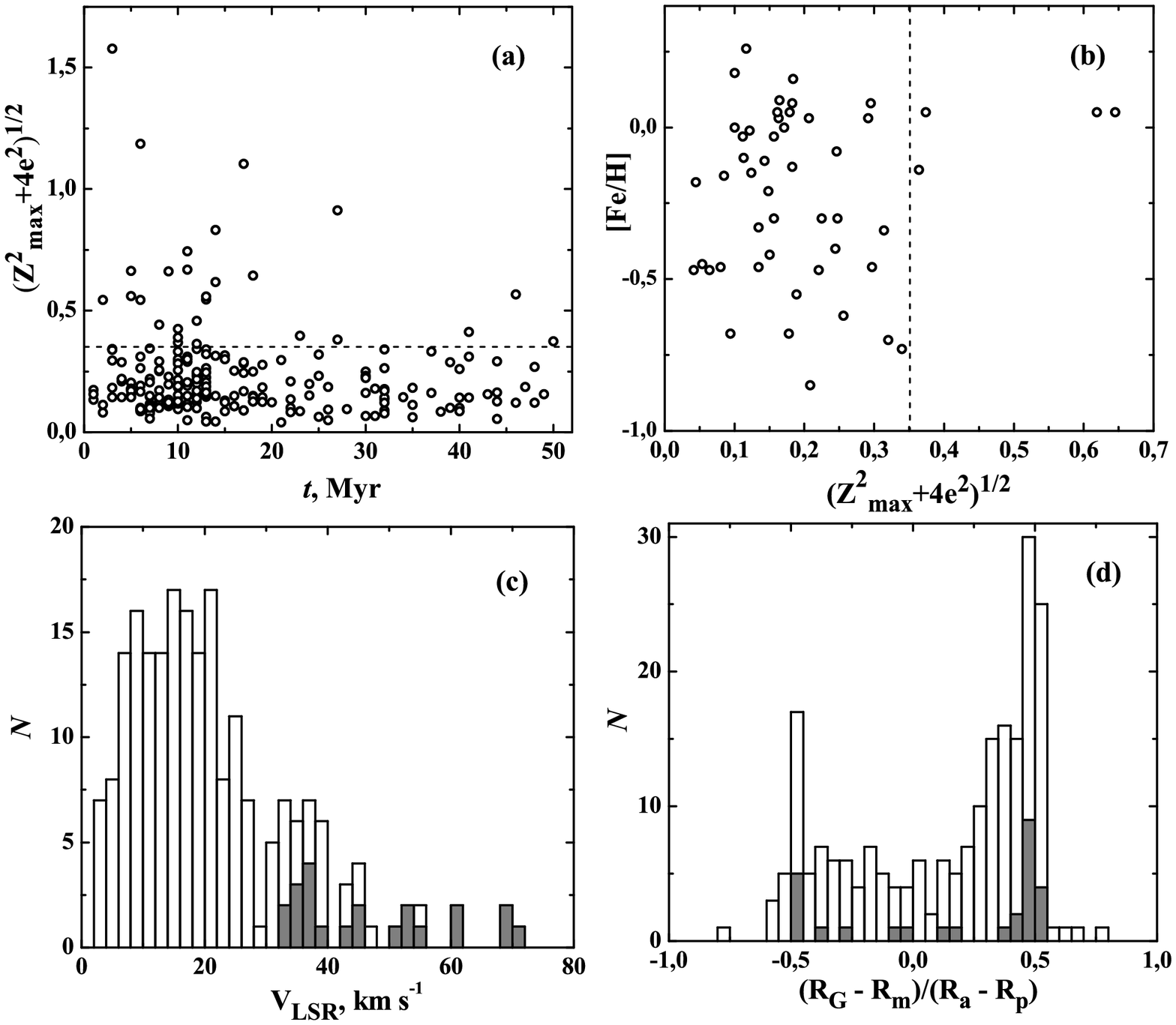}
\caption{(a) Age--–$(Z^2_{max} + 4e^2)^{1/2}$ diagram, 
         where $Z_{max}$ is the
         maximum distance of the orbital points from the Galactic
         plane expressed in kiloparsecs and $e$ is the orbital 
         eccentricity; (b) $(Z^2_{max} + 4e^2)^{1/2}$-–[Fe/H] 
         diagram; (c) the 
         distribution in total residual velocities; (d) the 
         distribution in relative distances from the mean radii 
         of their orbits $(R_{G} - R_{m})/(R_{a} - R_{p})$ 
         for open clusters younger 
         than 50~Myr. The dashed lines on the first two panels 
         correspond to the critical value of the parameter
         $(Z^2_{max} + 4e^2)^{1/2} = 0.35$; the gray histograms 
         on panels (c, d) 
         indicate the clusters for which this parameter $> 0.35$.}
\label{fig1}
\end{figure*}

\newpage

\begin{figure*}
\centering
\includegraphics[angle=0,width=0.90\textwidth,clip]{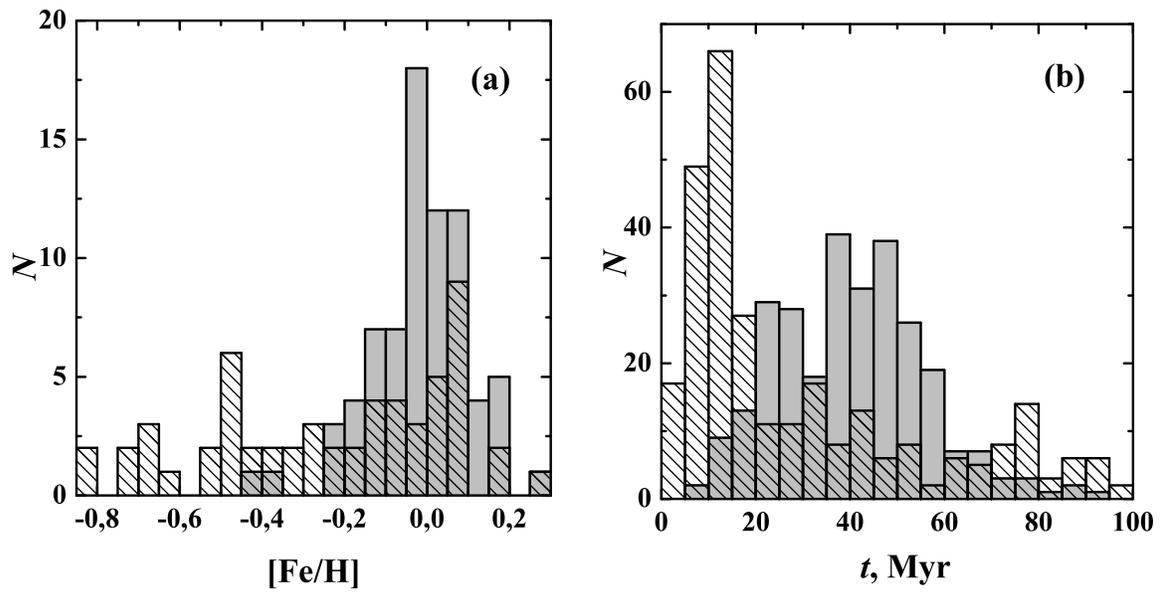}
\caption{The distributions of young ($t < 50$~Myr) open clusters 
         (hatched) and field Cepheids (gray histograms) in metallicity 
         (a) and age (b).}
\label{fig2}
\end{figure*}

\newpage

\begin{figure*}
\centering
\includegraphics[angle=0,width=0.56\textwidth,clip]{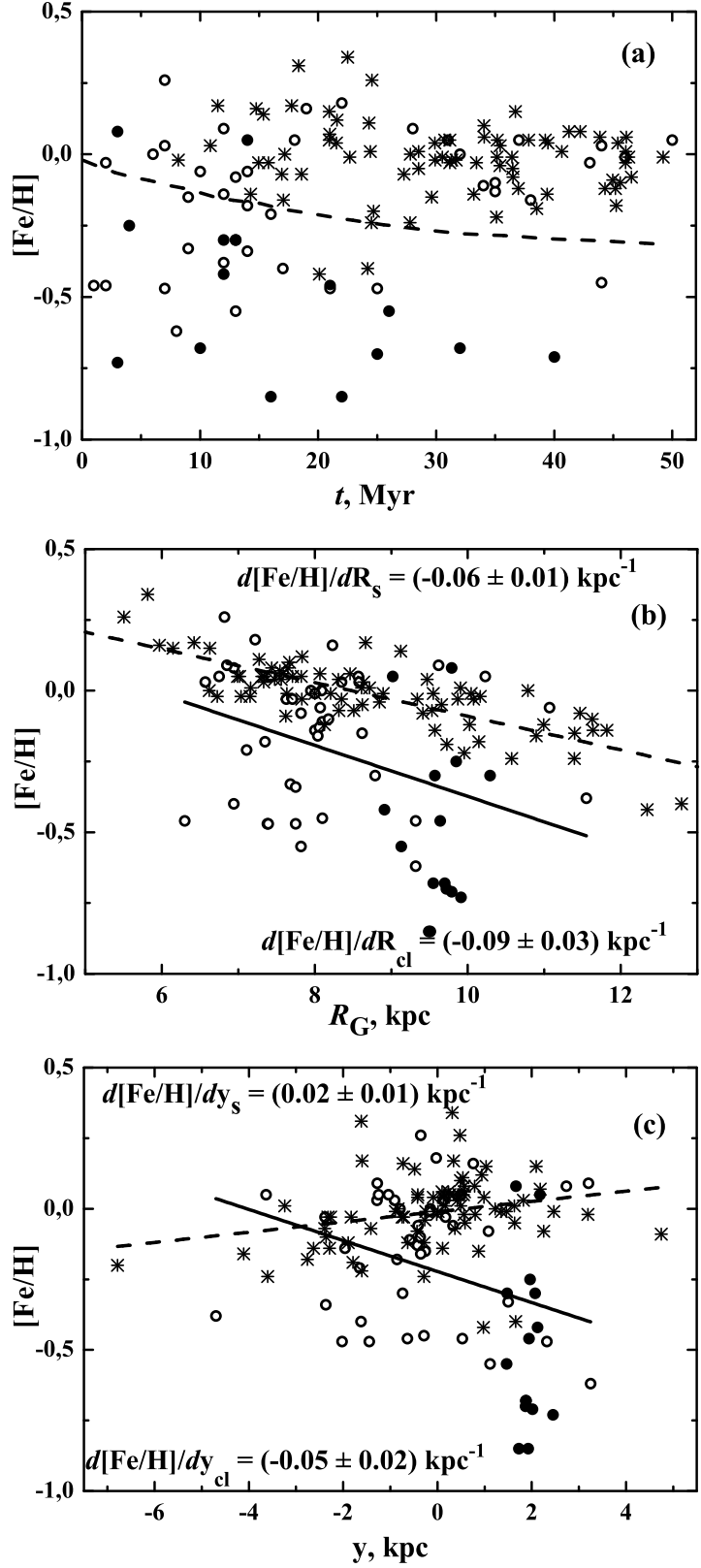}
\caption{Metallicity for young ($t < 50$ Myr) open clusters and field 
         Cepheids versus age (a), Galactocentric distance (b), and
         distance along the y coordinate (azimuthal distance). The 
         circles are clusters, the filled circles are clusters of 
         the Perseus complex, and the snowflakes are Cepheids. The 
         dashed curve on panel (a) is the lower envelope for field 
         Cepheids drawn by eye. The lines on panels (b) and (c) are 
         the linear regressions for clusters (solid) and Cepheids 
         (dashed). The radial and azimuthal metallicity gradients 
         are shown on the corresponding panels.}
\label{fig3}
\end{figure*}

\newpage

\begin{figure*}
\centering
\includegraphics[angle=0,width=0.96\textwidth,clip]{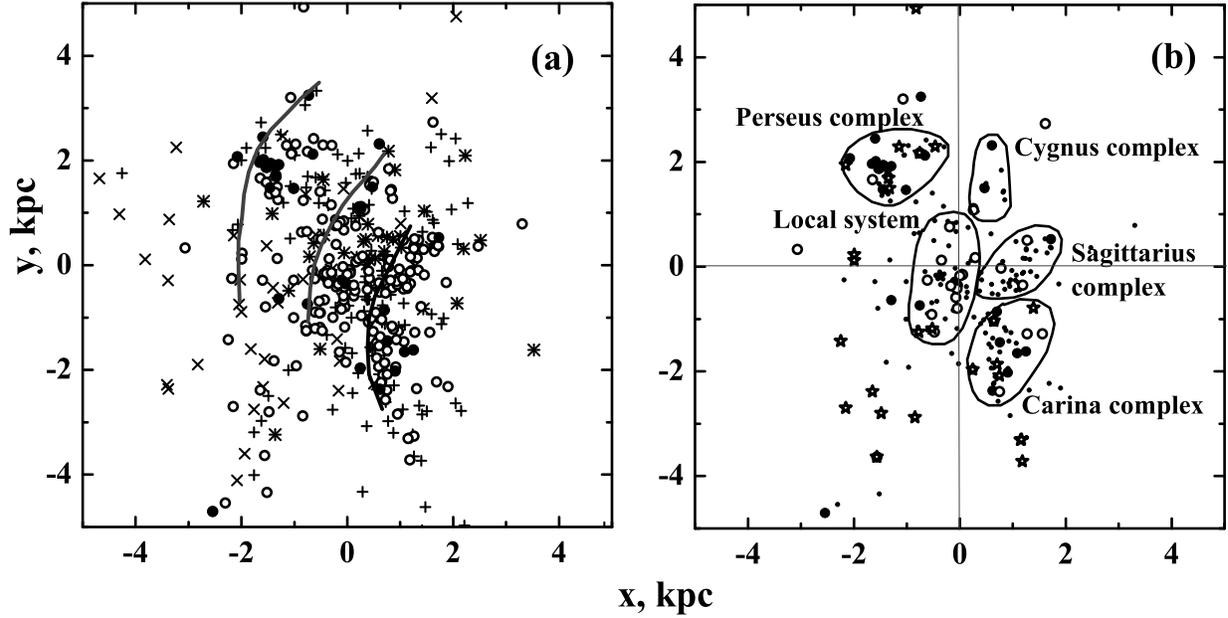}
\caption{(a): The distribution of young clusters and Cepheids in 
          projection onto the Galactic plane, where the circles 
          are clusters, the filled circles are metal-–poor clusters 
          lying below the dashed line in Fig.~3a, the crosses are 
          metal--poor Cepheids with $[Fe/H]< 0.0$, the snowflakes 
          are metal--rich Cepheids, and the pluses are Cepheids with 
          unknown metallicities. The nearly vertical curves drawn by 
          eye are the left envelopes of the regions with an enhanced 
          cluster density, while the middle and right curves are 
          simultaneously the right envelopes for the regions with an 
          enhanced density of field Cepheids. (b) The same for 
          clusters, the closed circles are metal--poor clusters, 
          the open circles are metal-rich clusters, the asterisks are 
          fast clusters, and the dots are clusters without metallicity. 
          The ovals drawn by eye are places of enhanced cluster 
          density-—the stellar complexes with their names.}
\label{fig4}
\end{figure*}

\newpage

\begin{figure*}
\centering
\includegraphics[angle=0,width=0.96\textwidth,clip]{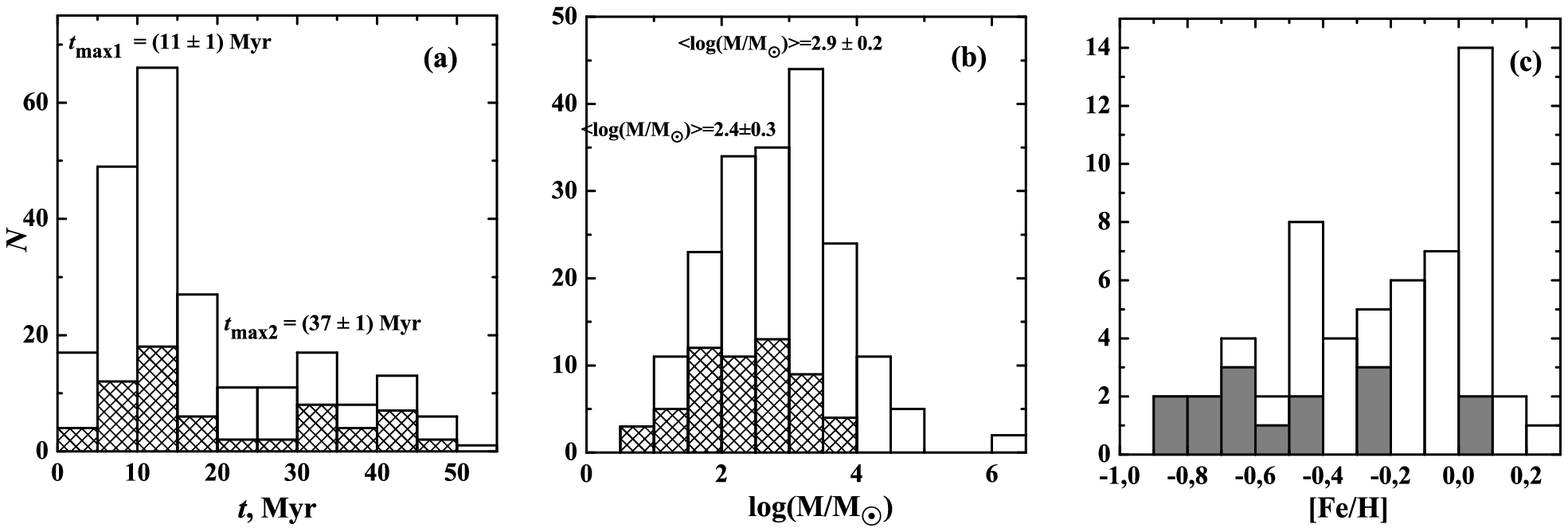}
\caption{The distributions of all young open clusters and only 
         Local--system clusters (hatched histograms) in age (a) 
         and mass (b); (c) the metallicity functions for all young 
         clusters and only for Perseus--complex clusters (gray histogram).}
\label{fig5}
\end{figure*}


\begin{thebibliography}{}

\bibitem{Andrievsky02a}
S.M.~Andrievsky, V.V.~Kovtyukh, R.E.~Luck, 
et\,al., Astron. Astrophys. {\bf 381}, 32 (2002а).

\bibitem{Andrievsky02b}
S.M.~Andrievsky, D.~Bersier, V.V.~Kovtyukh, 
et\,al., Astron. Astrophys. {\bf 384}, 140 (2002\,b).

\bibitem{Andrievsky02c}
S.M.~Andrievsky, V.V.~Kovtyukh, R.E.~Luck, 
et\,al., Astron. Astrophys. {\bf 392}, 491 (2002\,c).

\bibitem{Andrievsky04}
S.M.~Andrievsky, R.E.~Luck, P.~Martin, 
et\,al., Astron. Astrophys. {\bf 413}, 159 (2004).
 
\bibitem{Berdnikov03}
L.N.~Berdnikov, A.S.~Rastorguev,
A.K.~Dambis, et\,al., A Catalogue of
Data on Galactic Cepheids (2003), http://
www.sai.msu.su/groups/cluster/CEP/RADVEL/.

\bibitem{Chiappini97}
C.~Chiappini, F.~Matteucci, R.G.~Gratton, 
Astrophys. J. {\bf 477}, 765 (1997).

\bibitem{Gozha12}
M.L.~Gozha, T.V.~Borkova  and V.A.~Marsakov, Astron. 
Lett. {\bf 38}, 506 (2012a).

\bibitem{Gozha12}
M.L.~Gozha, V.V.~Koval'  and V.A.~Marsakov, Astron. 
Lett. {\bf 38}, 519 (2012b).

\bibitem{Efremov11}
Yu.N.~Efremov, Astron.Rep. {\bf 55}, 108 (2011). 

\bibitem{Ivanov83}
G.R.~Ivanov, Sov. Astron. Lett. {\bf 9}, 107 (1983). 

\bibitem{Kovtyukh05}
V.V.~Kovtyukh, G.~Wallerstein, and S.M.~Andrievsky,
Publ. Astron. Soc. Pacif. {\bf 117}, 1173 (2005).


\bibitem{Lepine11}
J.R.D.~Lepine, P.~Cruz, S.~Scarano Jr., 
et\,al., MNRAS {\bf 417}, 698 (2011). 


\bibitem{Luck03}
R.E.~Luck, W.P.~Gieren, S.M.~Andrievsky, 
et\,al., Astron. Astrophys. {\bf 401}, 939 (2003).

\bibitem{Putte10}
D.~Vande Putte, T.P.~Garnier, I. Ferreras, et\,al., 
Mon. Not. R. Astron. Soc. {\bf 407}, 2109 (2010).




\end{thebibliography}
\end{document}